\documentclass[11pt]{article}
\usepackage{graphicx}
\usepackage{epsfig}
\usepackage[figuresright]{rotating}
\usepackage{amsmath}
\usepackage{amssymb}
\usepackage{amsfonts}

\newcommand{\pibf}{\mbox{\boldmath $\pi$}}

\usepackage{cite}
\begin{document}
\begin{center}
{\LARGE{\bf Electromagnetic  polarizabilities and the excited states of the
    nucleon }}\\[1ex]
Martin Schumacher\\mschuma3@gwdg.de\\
Zweites Physikalisches Institut der Universit\"at G\"ottingen,
Friedrich-Hund-Platz 1\\ D-37077 G\"ottingen, Germany
\end{center}

\begin{abstract}
The electromagnetic polarizabilities of the nucleon are shown to be
essentially composed of the nonresonant  $\alpha_p(E_{0+})=+3.2$,  
 $\alpha_n(E_{0+})=+4.1$, the $t$-channel
$\alpha^t_{p,n}=-\beta^t_{p,n}=+7.6$ and the resonant
$\beta_{p,n}(P_{33}(1232))=+8.3$ contributions
(in units of $10^{-4}{\rm fm}^3$). The remaining deviations from the
experimental data $\Delta\alpha_p=1.2\pm 0.6$,  
$\Delta\beta_p=1.2\mp 0.6$,
$\Delta\alpha_n=0.8\pm 1.7$ and 
$\Delta\beta_n=2.0\mp 1.8$ are contributed by a larger number of resonant and 
nonresonant processes with cancellations between the contributions. This
result confirms that dominant contributions to the electric and magnetic
polarizabilities may be represented in terms of  two-photon 
coupling to the $\sigma$ meson having
the predicted  mass $m_\sigma = 666$ MeV and two-photon width 
$\Gamma_{\gamma\gamma}
= 2.6$ keV.
\end{abstract}

{\bf PACS.} 11.55.Fv Dispersion relations -- 11.55.Hx Sum rules --
13.60.Fz Elastic and Compton scattering -- 14.70.Bh Photons 

\section{Introduction}
Recently, it has been shown \cite{schumacher06} 
that the $t$-channel components $\alpha^t$ and 
$\beta^t$ of the electric ($\alpha$) and magnetic ($\beta$) polarizabilities
of the nucleon can be understood as a property of the constituent quarks.
The constituent quarks couple to   $\pibf$ and $\sigma$ fields and, mediated
by these fields, they couple  to two photons. The coupling of two photons with
perpendicular linear polarization to the $\pi^0$ meson 
provides the main contribution, $\gamma^t_\pi$, to  the backward-angle
spin-polarizability $\gamma_\pi$. Similarly, the coupling of two photons
with parallel linear polarization provides the main contribution,
$(\alpha-\beta)^t$, to the difference ($\alpha-\beta)$
of the electric and the magnetic polarizabilities. The quantitative
prediction $(\alpha-\beta)^t_{p,n} = 15.2$ (in units of $10^{-4}{\rm fm}^3$) 
makes use of the fact that the mass of the particle of the $\sigma$ field is   
predicted by the quark-level Nambu--Jona-Lasinio (NJL) model 
to be $m_\sigma=666$ MeV and its  two-photon width to be 
$\Gamma_{\gamma\gamma}=2.6$ keV \cite{schumacher06}. 

The foregoing paragraph describes the r\'esum\'e of a long and partial
controversial history of research. The scalar-isoscalar $t$-channel was
introduced \cite{hearn62} in analogy to the pseudoscalar $t$-channel 
\cite{low60}. But differing from the $\pi^0$-pole contribution \cite{low60}
to the scattering  amplitude, the meaning and importance 
of the scalar-isoscalar $t$-channel \cite{hearn62} was less well known,
mainly  because the $\sigma$ meson was not considered as a normal particle.
One important step forward was the formulation of the BEFT sum rule  
\cite{bernabeu74}, relating the $s$-channel part of the difference of the
electric and magnetic polarizabilities, $(\alpha-\beta)^s$, to the multipole
content of the total photoabsorption cross section using a fixed-$\theta$
dispersion relation at $\theta=\pi$, and by relating the $t$-channel part
$(\alpha-\beta)^t$ to a dispersion relation for $t$ with the imaginary part
of the amplitude taken from the reactions $\gamma\gamma\to\pi\pi$ and
$N\bar{N}\to\pi\pi$ via a unitarity relation. Furthermore, the scalar-isoscalar
phase $\delta^0_0(t)$ was taken from the reaction $\pi\pi\to\pi\pi$.
One of the first evaluations of the BEFT sum-rule showed that for pointlike
uncorrelated  pions the large
value of $(\alpha-\beta)^t=+17.51$  is
obtained \cite{guiasu76} which in other calculations has been 
reduced by very different  factors (see
\cite{schumacher05b} for an overview)
 when the $\pi\pi$ correlation and the pion internal structure is 
taken into account.  The largest 
reduction amounting to  a factor of  2 has been obtained in the latest of this 
early series of calculations \cite{holstein94}.
This unsatisfactory situation has recently been clarified
by showing \cite{schumacher06,levchuk05} that the arithmetic average
of the most recent calculations of Drechsel et al. \cite{drechsel03}
and Levchuk et al. (see \cite{schumacher05b}),
$(\alpha-\beta)^t_{p,n}= 15.3\pm 1.3$, leads to a very good agreement with 
the experimental result and with a parameter-free calculation based on the
quark-level NJL model or dynamical linear $\sigma$ model 
(L$\sigma$M) \cite{schumacher06,hatsuda94,delbourgo95}, leading to 
$(\alpha-\beta)^t_{p,n}= 15.2$.

After the size and the dynamics of the $t$-channel contribution to the
electromagnetic polarizabilities has been well  understood, it appears of
interest to get a similar understanding for the $s$-channel contribution.
Especially, the question has to be answered what the individual
contributions of 
the resonant excited states of the nucleon to the electric and magnetic
polarizabilities are and how the contributions of the ``pion cloud'' 
to the electric and magnetic polarizabilities may  be
specified. To the author's knowledge such an investigation has not been carried
out before.

\section{Electromagnetic polarizabilities obtained from the forward-angle 
sum-rule for $(\alpha+\beta)$ and the backward-angle  sum-rule for
$(\alpha-\beta)$}

The appropriate tool for the present investigation is to simultaneously apply 
the  forward-angle 
sum-rule for $(\alpha+\beta)$ and the backward-angle  sum-rule for
$(\alpha-\beta)$. This leads to the following relations

\begin{eqnarray}
&& \alpha=\alpha^s+ \alpha^t, \label{alph-1}\\
&&\alpha^s=\frac{1}{2\pi^2} \int^\infty_{\omega_0} \left[
A(\omega)\,\sigma(\omega,E1,M2,\cdots)
+B(\omega)\,\sigma(\omega,M1,E2,\cdots)\right]
\frac{d\omega}{\omega^2}, \label{alph-2}\\
&&\alpha^t=\frac{5 \, \alpha_e \, g_{\pi NN}}{12\,\pi^2\,m^2_\sigma\,f_\pi}
=7.6,
\label{alph-3}
\end{eqnarray}
and 
\begin{eqnarray}
&&\beta=\beta^s+\beta^t, \label{beta-1}\\
&&\beta^s=\frac{1}{2\pi^2}\int^\infty_{\omega_0}\left[
A(\omega)\,\sigma(\omega,M1,E2,\cdots)
+B(\omega)\,\sigma(\omega,E1,M2,\cdots)\right]
\frac{d\omega}{\omega^2}, \label{beta-2} \\
&&\beta^t=-\frac{5 \, \alpha_e \, g_{\pi NN}}{12\,\pi^2\,m^2_\sigma\,f_\pi}
=-7.6, \label{alph-4}\\
&&\omega_0=m_\pi+\frac{m^2_\pi}{2\,m}, \label{threshold}\\
&&A(\omega)=\frac12 \left(1+\sqrt{1+\frac{2\omega}{m}}\right), \label{A}\\
&&B(\omega)=\frac12 \left(1-\sqrt{1+\frac{2\omega}{m}}\right). \label{B}
\end{eqnarray}
In (\ref{alph-1}) --
(\ref{B}) $\omega$ is the photon energy in the lab-system, $m_\pi$
the pion mass and $m$ the nucleon mass.  
The quantities  $\alpha^s$, $\beta^s$
are the $s$-channel electric and magnetic polarizabilities, and 
$\alpha^t$, $\beta^t$ the $t$-channel electric and magnetic polarizabilities,
respectively. The multipole content of the photoabsorption
 cross section enters through
\begin{eqnarray}
&&\sigma(\omega,E1,M2,\cdots)=\sigma(\omega,E1)+\sigma(\omega,M2)+\cdots,
\label{cross1}\\
&&\sigma(\omega,M1,E2,\cdots)=\sigma(\omega,M1)+\sigma(\omega,E2)+\cdots,
\label{cross2}
\end{eqnarray}
{\it i.e.} through the sums of cross sections with change and without 
change of parity during the electromagnetic transition, 
respectively\footnote{It should be noted that this separation into cross 
sections for separate multipoles is possible in the presently used 
fixed-$\theta$ dispersion theory applied at $\theta=0$ and 
$\theta=\pi$, whereas in the corresponding formulas
based on fixed-$t$ dispersion theory \cite{lvov97}
terms containing mixed products of 
CGLN \cite{chew57}
amplitudes occur (for a discussion see \cite{schumacher05b}).}.
The multipoles belonging to parity-change are favored for the electric
polarizability $\alpha^s$ whereas the multipoles belonging to parity-nonchange
are favored for the magnetic polarizability $\beta^s$.   
The coefficients $A(\omega)$ and $B(\omega)$ in Eqs. (\ref{alph-2}),
(\ref{beta-2}), (\ref{A}) and (\ref{B})  multiplying the cross 
sections of the
parity-favored  and parity-nonfavored  multipoles, respectively,
are  $A \sim + 1.07$ and $B \sim -0.07$ at the pion photoproduction
threshold. They increase with photon energy, as expected for relativistic
correction factors. Using $A(\omega)$ and $B(\omega)$
it is easy to prove that $(\alpha+\beta)\equiv (\alpha+\beta)^s$ is given by
the Baldin or Baldin-Lapidus (BL) \cite{baldin60}
sum rule, whereas $(\alpha-\beta)^s$
is given by the $s$-channel part of the  BEFT \cite{bernabeu74} sum rule.

For the $t$-channel parts, $\alpha^t$ and $\beta^t$, we use the 
predictions obtained from the $\sigma$-meson pole 
representation\footnote{This $\sigma$-meson pole in the 
complex $t$-plane of the Compton
scattering amplitude $A(s,t)$ is not the same,
but has relations with the $\sigma$-meson pole introduced to parameterize the 
$\pi\pi$ scattering amplitude. These relations have been discussed in detail
in \cite{schumacher06,schumacher05b}.}
with properties as predicted
by the quark-level Nambu--Jona-Lasinio model \cite{levchuk05,schumacher06}. 
The quantities entering into this prediction are 
$\alpha_e=e^2/4\pi=1/137.04$,   
the pion-nucleon coupling constant,
$g_{\pi NN}= 13.169\pm 0.057$, 
the pion decay constant, $f_\pi=(92.42\pm 0.26)$ MeV, 
and the $\sigma$-meson mass, $m_\sigma= 666.0$ MeV
\cite{schumacher06,hatsuda94,delbourgo95}. For convenience we summarize the
arguments leading to the relations (\ref{alph-3}) and (\ref{alph-4}).
The flavor wave-functions of the $\pi^0$ and the $\sigma$ meson are given by
\begin{equation}
|\pi^0\rangle=\frac{1}{\sqrt{2}}(-u\bar{u}+d\bar{d}),\quad 
|\sigma\rangle=\frac{1}{\sqrt{2}}(u\bar{u}+d\bar{d}).
\label{flavorwf}
\end{equation} 
This leads to the decay matrix elements
\begin{equation}
M(\sigma\to\gamma\gamma)= -\frac53
M(\pi^0\to\gamma\gamma)=\frac53\frac{\alpha_e}{\pi f_\pi}. 
\label{matrix}
\end{equation}
Using the NJL model or the dynamical L$\sigma$M with dimensional regularization
we arrive at \cite{schumacher06,delbourgo95}
\begin{equation}
m^{\rm cl}_\sigma=\frac{4\pi f^{\rm cl}_\pi}{\sqrt{N_c}},
\label{dimensional}
\end{equation}
where $m^{\rm cl}_\sigma$ and $f^{\rm cl}_\pi=89.8$ MeV are the 
$\sigma$ meson mass and the $\pi$ decay constant in the chiral limit (cl) and
$N_c=3$ the number of colors.
Then the mass of the $\sigma$ meson is given by
\begin{equation}
m_\sigma=\sqrt{(m^{\rm cl}_\sigma)^2+m^2_\pi}= 666\,\,\, \mbox{MeV}.
\label{sigmamass}
\end{equation}
Inserting this into 
\begin{equation}
(\alpha-\beta)^t=\frac{g_{\sigma NN}M(\sigma\to\gamma\gamma)}{2\pi m^2_\sigma}
\label{sigmat}
\end{equation}
and using $f_{\sigma NN}=f_{\pi NN}$ and $(\alpha+\beta)^t=0$ 
we arrive at (\ref{alph-3}) and (\ref{alph-4}).

\section{Components of electromagnetic polarizabilities from analyses
of total photoabsorption and meson photoproduction data}

In the following we use different photoabsorption data to get information
on partial contributions to $\alpha^s$ and $\beta^s$. Analyses of total
photoabsorption cross sections have been carried out in \cite{armstrong72}.
These analyses give a very good overview over the resonant and nonresonant
contributions to the electromagnetic polarizabilities. Further information
is taken from the PDG2006 \cite{yao06}, the GWSES \cite{arndt02} and the 
Mainz \cite{hanstein98,drechsel99} analyses of meson photoproduction data.

\subsection{Components of electromagnetic polarizabilities from analysis
of the total photoabsorption cross-section of the proton}

In the following we wish to study the contributions of nucleon
resonances and nonresonant excited states
to the $s$-channel electromagnetic polarizabilities. Only the 
resonances  $P_{33}(1232)$, $P_{11}(1440)$,  
$D_{13}(1520)$,  $S_{11}(1535)$ and $F_{15}(1680)$
have to be taken into account. The contributions of the resonances
$S_{11}(1650)$, $D_{15}(1675)$ and higher lying resonances are negligible.
For this analysis we use the Walker \cite{walker69,armstrong72}
parameterization of nucleon resonances
\begin{eqnarray}
&&I=I_r\left(\frac{ k_r}{k}\right)^2
\frac{W^2_r\,\Gamma\,\Gamma_\gamma}{
(W^2-W^2_r)^2+W^2_r\,\Gamma^2},\label{arm1}\\
&&\Gamma=\Gamma_r\left(\frac{q}{q_r}\right)^{2l+1}
\left(\frac{q^2_r+X^2}{
q^2+X^2}\right)^l, \label{arm2}\\
&&\Gamma_\gamma=\Gamma_r\left(\frac{k}{k_r}\right)^{2j_\gamma}\left(\frac{
k^2_r+X^2}{k^2+X^2}\right)^{j_\gamma}.\label{arm3}
\end{eqnarray}
\begin{eqnarray}
&&s=2\omega m+m^2,\quad \mbox{$\omega$ =  photon energy in the lab system,}\\
&&W^2=s\\
&&k=|{\bf k}|=\frac{s-m^2}{2\sqrt{s}},\quad\mbox{$|{\bf k}|$ = photon momentum 
in the c.m. system}, \\
&&q=|{\bf q }|=\sqrt{E^2_\pi-m^2_\pi};\,E_\pi=\frac{s-m^2+ m^2_\pi}{2\sqrt{s}},
\mbox{$|{\bf q }|$ = $\pi$ momentum in the c.m. system},\\
&&j_\gamma,\quad\mbox{multipole angular momentum of the photon},\\
&& l,\quad \mbox{single $\pi$ angular momentum}.
\end{eqnarray}
The damping constants $X$ are $X=160$ MeV for the $P_{33}(1232)$ resonance
and  $X=350$ MeV else.

For the proton,  parameters are given in \cite{armstrong72} for the relevant
resonant states, leading to the results given in lines 3 -- 5 of Table
\ref{armstrong}. 
\begin{table}[h]
\caption{Partial contributions to the electromagnetic polarizabilities based
  on the analysis of the total photoabsorption cross section 
\cite{armstrong72}.     The $t$-channel parts in line 6 are the predictions
based on the $\sigma$-meson pole representation (see section 2). 
Line 7 contains the differences between the
numbers in line 2 and the sums of numbers given in lines 3--6. The
experimental data are normalized to $(\alpha+\beta)_p=13.9 \pm 0.3$ (see 
\cite{schumacher05b}).}
\begin{center}
\begin{tabular}{ll|ll|}
\hline
1& &$\alpha_p$   &   $\beta_p$\\
\hline
2 &experiment & $12.0\pm 0.6$& $1.9 \mp 0.6$\\
\hline
3& $P_{33}(1232)$\, $M1,E2$ & $ -1.1$  & $+8.3$ \\
4& $P_{11}(1440)$\, $M1$ & $ -0.1$  & $+0.3$ \\
5& $D_{13}(1520)$\, $E1,M2$ & $+1.2$      &  $-0.3$\\ 
6& $S_{11}(1535)$\, $E1$ & $+0.1$      &  $-0.0$\\ 
5& $F_{15}(1680)$\, $E2,M3$ & $-0.1$   & $+0.4$\\ 
6& $t$-channel& $+7.6$ & $-7.6$\\
7& nonresonant & $+4.4$ & $+0.8$ \\
\hline
\label{armstrong}
\end{tabular}
\end{center}
\end{table}
The sum $\alpha_p + \beta_p$ of nonresonant contributions in line 7 of 
Table \ref{armstrong} is in agreement with   the 
corresponding number
calculated from the nonresonant cross section given in   \cite{armstrong72}
if the nonresonant cross section data are extrapolated to about 3.5 GeV. 
This shows
that with the predicted $t$-channel contributions given in line 6 there is 
consistency between the experimental electromagnetic polarizabilities 
and the predictions.

\subsection{Components of electromagnetic polarizabilities from analyses
of meson photoproduction  for  the proton and the neutron}

From isospin considerations it has been derived \cite{watson54} that 
the amplitudes for meson photoproduction are composed of 
$A^{(1/2)}$ and $A^{(3/2)}$, referring to final states of definite
isospin ($\frac12$ or $\frac32$). Furthermore, there is an amplitude $A^{(0)}$
which may be related to ``recoil'' effects \cite{watson54}.
This latter amplitude makes a contribution to $I=1/2$ only. Therefore,
the amplitudes 
\begin{equation}
_pA^{(1/2)}=A^{(0)}+\frac13
A^{(1/2)}, \quad  _nA^{(1/2)}=A^{(0)}-\frac13 A^{(1/2)}
\label{pnAmplitudes}
\end{equation}
may be introduced. 
Furthermore, with
\begin{equation}
A^{(+)}=\frac13 (A^{(1/2)}+2 A^{(3/2)}),\quad
A^{(-)}=\frac13 (A^{(1/2)}- A^{(3/2)}), \\
\label{+-Amplitudes}
\end{equation}
the physical amplitudes may be expressed by the isospin combinations
(see e.g. \cite{drechsel99,ericson88})
\begin{eqnarray}
&&A(\gamma p\to n \pi^+)=\sqrt{2}(A^{(-)}+A^{(0)})
=\sqrt{2}(_pA^{(1/2)}-\frac13 A^{(3/2)}),\\ 
&&A(\gamma p\to p \pi^0)=A^{(+)}+A^{(0)}=_pA^{(1/2)}+\frac23
  A^{(3/2)},\\ 
&&A(\gamma n\to p \pi^-)=-\sqrt{2}(A^{(-)}-A^{(0)})=
\sqrt{2}(_nA^{(1/2)}+\frac13 A^{(3/2)}),\\ 
&&A(\gamma n\to n \pi^0)=A^{(+)}-A^{(0)}=- _nA^{(1/2)}+\frac23
  A^{(3/2)}.
\label{amplitudes1}
\end{eqnarray}

The relation for  the cross section of 1$\pi$ photoproduction is given by
\begin{eqnarray}
&&\sigma^{1\pi}=2\pi\frac{|\bf q|}{|\bf k|}\sum^\infty_{l=0}(l+1)^2
\left[(l+2)(|E_{l+}|^2+|M_{(l+1)-}|^2)+l(|M_{l+}|^2+|E_{(l+1)-}|^2)\right],
\label{sigtot1}\\
&&\Delta\sigma^{1\pi}=2\pi\frac{|\bf q|}{|\bf k|}\sum^\infty_{l=0}(l+1)^2(-1)^l
\left[(l+2)(|E_{l+}|^2-|M_{(l+1)-}|^2)+l(|M_{l+}|^2-|E_{(l+1)-}|^2)\right],
\label{sigtot2}\\
&&\Delta\sigma^{1\pi}=\sigma^{1\pi}(E1,M2,\cdots)-\sigma^{1\pi}(M1,E2,\cdots).
\label{sigtot3}
\end{eqnarray}

The peak cross section $I_r$ introduced in (\ref{arm1}) is given by
\begin{equation}
I_r=2\pi\,\frac{1}{k^2_r}\,\frac{2J+1}{2J_0+1}\,\frac{\Gamma_\gamma}{\Gamma},
\label{peakcross}
\end{equation}
where $J$ and $J_0$ are the spins of the excited state and the ground state,
respectively, $\Gamma_\gamma$ the photon width and $\Gamma$ the total width
of the resonance. The photon  width $\Gamma_\gamma$ may be expressed
through the resonance couplings $A_{1/2}$ and $A_{3/2}$ by the relation
\cite{yao06}
\begin{equation}
\Gamma_\gamma=\frac{k^2_r}{\pi}\,\frac{2\,M_N}{(2J+1)M_R}\left[|A_{1/2}|^2 +
  |A_{3/2}|^2\right], \label{resoncoup1}
\end{equation} 
where $M_N$ and $M_R$ are the nucleon and resonant
masses.  Combining
(\ref{peakcross}) and (\ref{resoncoup1})
we arrive at\footnote{It should be noted that the quantity $I_r$ 
of  (\ref{IRformula}) contains the branching correction $\Gamma/\Gamma_\pi$
as required.}
\begin{equation}
I_r=\frac{2\,M_N}{M_R \,\Gamma}\left[|A_{1/2}|^2+|A_{3/2}|^2\right].
\label{IRformula}
\end{equation}
Using  (\ref{IRformula}) the quantity $I_r$  can  be calculated from 
the resonance couplings $A_{1/2}$ and $A_{3/2}$ given 
by the PDG \cite{yao06}, by 
GWSES \cite{arndt02} and Mainz \cite{drechsel99}. The results obtained
for the electromagnetic polarizabilities obtained from the data given
in \cite{drechsel99} are given in lines 3 -- 7 of Table \ref{drechsel}.
\begin{table}[h]
\caption{Partial contributions to the electromagnetic polarizabilities. The
  resonant contributions in lines 3--7 are obtained from the analysis 
of Drechsel  et al. \cite{drechsel99}.
The $t$-channel parts in line 8 are  the predictions based on the 
$\sigma$-meson pole representation (see section 2).
The predicted contribution due to the $E_{0+}$ amplitude in line 9
is based on the analyses given in \cite{arndt02,hanstein 98,drechsel99}. 
Line 10 contains the
differences between the numbers in line 2 and the sums of numbers given in
lines 3--9. The experimental data are normalized to
$(\alpha+\beta)_p=13.9\pm 0.3$ and $(\alpha+\beta)_n=15.2\pm 0.5$ (see
  \cite{schumacher05b}).}
\begin{center}
\begin{tabular}{ll|ll|ll|}
\hline
1& &$\alpha_p$   &   $\beta_p$    &   $\alpha_n$ &   $\beta_n$\\
\hline
2 &experiment & $12.0\pm 0.6$& $1.9 \mp 0.6$& $12.5\pm 1.7$& $2.7\mp 1.8$\\
\hline
3& $P_{33}(1232)$\, $M^{(3/2)}_{1+},E^{(3/2)}_{1+} $ & 
$ -1.1$  &  $+8.3$  &   $-1.1$    &   $+8.3$\\
4& $P_{11}(1440)$\, $_{p,n}M^{(1/2)}_{1-}$ & 
$ -0.0$  &  $+0.2$  &   $-0.0$    &   $+0.1$\\
5& $D_{13}(1520)$\, $_{p,n}E^{(1/2)}_{2-},{} _{p,n}M^{(1/2)}_{2-}$ 
& $+0.6$ &  $-0.2$  &   $+0.5$    &   $-0.1$\\  
6& $S_{11}(1535)$\, $_{p,n}E^{(1/2)}_{0+}$ & $+0.1$      
&  $-0.0$  &   $+0.1$    &   $-0.0$\\ 
7& $F_{15}(1680)$\, $_{p,n}E^{(1/2)}_{3-}, {}_{p,n}M^{(1/2)}_{3-}$ 
& $-0.1$   & $+0.3$  & $-0.0$   &   $+0.0$ \\ 
8& $t$-channel& $+7.6$ & $-7.6$& $+7.6$& $-7.6$\\
9& $E_{0+}$ (empirical)& $+3.2$ & $-0.3$ & $+4.1$& $-0.4$\\
10& background & $+1.7$&$+1.2$ & $+1.3$& $+2.4$\\
\hline
\label{drechsel}
\end{tabular}
\end{center}
\end{table}

The main contributions to the nonresonant parts of the electromagnetic
polarizabilities are expected from the $E_{0+}$ amplitude which has to be
taken from analyses of meson photoproduction data. Multipole analyses of pion
photoproduction based on fixed-$t$ dispersion relations and unitarity
are given  by Hanstein et al. \cite{hanstein98} in a convenient form.
Cross sections separated into resonant and nonresonant parts are provided 
for the reactions $\gamma p \to \pi^+ n$
and $\gamma n \to \pi^- p$ up to energies of 500 MeV and 
extrapolations of the nonresonant parts are straightforward  using the data
contained in \cite{drechsel99} and \cite{arndt02}. In principle there is a
problem in disentangling resonant and nonresonant contributions because
of interference effects. The interference of the amplitudes
$_{p,n}E^{(1/2)}_{0+}$ with the $S_{11}(1535)$ and $S_{11}(1650)$ resonances,
however, does not lead to problems in determining the nonresonant $E_{0+}$
contributions because of the smallness of the resonant parts.
The results for the electromagnetic polarizabilities
obtained from these empirical $E_{0+}$ data are contained in line 9 of Table
\ref{drechsel}.   
 
Up to this point the  electromagnetic polarizabilities find an 
explanation in the numbers given in lines 3 -- 9 of
Table \ref{drechsel}, with the exception of the small contributions 
given in line 10 which deserve a further investigation.
These non-$E_{0+}$ parts of the nonresonant contributions are 
partly  due to the $M^{(3/2)}_{1-}$, $_{p,n}M^{(1/2)}_{1+}$ 
and $_{p,n}E^{(1/2)}_{1+}$ amplitudes which interfere with the corresponding
resonant amplitudes $_{p,n}M^{(1/2)}_{1-}$ ($P_{11}(1440)$),  
$M^{(3/2)}_{1+}$ ($P_{33}(1232)$) and  $E^{(3/2)}_{1+}$ ($P_{33}(1232)$),
respectively \cite{drechsel99}.  
Only the nonresonant parts of the $M_{1-}$ and $M_{1+}$
amplitudes are expected to be  
to some extent important in comparison with dominant $E_{0+}$
amplitude. Therefore we restrict the present discussion to the $M_{1-}$
and $M_{1+}$
amplitudes. Using the data given in \cite{drechsel99} we arrive at the
estimates\\
$\alpha^{\rm nonres.}_p(M_{1-})=-0.0$, 
$\beta^{\rm nonres.}_p(M_{1-})=+0.2$, $\alpha^{\rm nonres.}_n(M_{1-})=-0.1$,
$\beta^{\rm nonres.}_p(M_{1-})=+0.4$, \\
$\alpha^{\rm nonres.}_p(M_{1+})=-0.0$, 
$\beta^{\rm nonres.}_p(M_{1+})=+0.3$, $\alpha^{\rm nonres.}_n(M_{1+})=-0.1$,
$\beta^{\rm nonres.}_p(M_{1+})=+0.6$.\\
The conclusion we have to draw from this is that it is not
possible to relate the numbers given in line 10 of Table \ref{drechsel}
to known photoproduction processes, unless the two-pion channels are taken
into account (see e.g. \cite{drechsel92}). The $\pi\pi N$ final states can be
characterized either as quasi two-body states such as $\pi\Delta$ and
$\rho N$, or as a $\pi\pi N$ component in which both pions are in $S$ waves. 
Furthermore, in the Regge regime above $\approx 2000$ MeV also 
$f_2(1270)$, $a_2(1320)$ and Pomeron $t$-channel exchanges play a role.
The  $\pi\Delta$ contribution has been analyzed in terms of a 
 $\Delta$ Kroll-Ruderman term and a 
$\Delta$ pion-pole term \cite{tejedor94}. Using data from this  analysis
\cite{tejedor94}
we arrive at $(\alpha_{p,n}+\beta_{p,n})\approx 1.0 $ for this partial
$\pi\pi$ channel. The nonresonant cross section above
$\approx 2000$ MeV makes a contribution of about 
 $(\alpha_{p,n}+\beta_{p,n})\approx 0.7 $.

\section{Discussion}
\subsection{Discussion of the $s$-channel contribution}

For a long time there have been  attempts to understand the electromagnetic
polarizabilities predominantly in terms of  properties of the 
``pion  cloud'' of the nucleon. Among these attempts CHPT in its original
relativistic form \cite{bernard91} 
is among the most prominent ones. It has been shown by L'vov
\cite{lvov93}
that the
results obtained for the electromagnetic polarizabilities through the
evaluation of chiral loops \cite{bernard91} can be reproduced via 
dispersion theory when 
the Born approximation of the electric-dipole CGLN amplitude $E_{0+}$
is  taken into account. 
The results obtained in this way  are shown in lines 2 and 3 of Table
\ref{table2}.
\begin{table}[h]
\caption{Predictions for the ``meson cloud''
 contribution to the electromagnetic
  polarizabilities in different approaches.}
\begin{center}
\begin{tabular}{ll|cc|cc|l}
\hline
1&method & $\alpha_p$ & $\beta_p$ & $\alpha_n$ &$\beta_n$ & reference\\
\hline
2&CHPT& $+7.4$ & $-2.0$ & $+10.1$ & $-1.2$ & Bernard \cite{bernard91}\\
3&pion$^{a)}$ Born & $+7.3$& $-1.8$ & $+9.8$ & $-0.9$ & L'vov \cite{lvov93}\\
4&$E0+$ Born& $+7.5$ & $-1.4$ & $+9.9$   & $-1.8$ & present \\ 
\hline
\end{tabular}
\end{center}
\label{table2}
a) The use of fixed-$t$ dispersion theory requires the consideration of
interference terms of the $E_{0+}$ amplitude with other amplitudes.
\end{table}

It is of interest to  use also the present approach based on forward and 
backward dispersion relations  for studies of this type. 
For this purpose use may be made of 
the Born approximation (see \cite{ericson88} p. 286, \cite{donnachie72}
p. 35) given in the form
\begin{eqnarray}
&&E^{\rm Born}_{0+}(\gamma N\to \pi^{\pm} N)= \pm \sqrt{2}\left( 
E^{(-){\rm Born}}_{0+} \pm E^{(0){\rm Born}}_{0+}\right),\\
&&E^{(-){\rm Born}}_{0+}= \frac{e f}{4\pi m_\pi}
\left[1-\frac12\left(1+\frac{1-v^2}{2v}\ln\left(\frac{1-v}{1+v}
\right)\right)\right], 
\label{E0+} 
\end{eqnarray}
with $v=|{\bf q}|/\sqrt{{\bf q}^2+ m^2_\pi}$ being the velocity of the pion
in the c.m. system. The expression given in (\ref{E0+}) corresponds to the
static approximation discussed in detail in \cite{ericson88,donnachie72}.
Because of the relation
\begin{equation}
\frac{\sigma_{E_{0+}}(\gamma n\to \pi^- p)}
{\sigma_{E_{0+}}(\gamma p\to \pi^+ n)}
=\left( 1+\frac{m_\pi}{m}\right)^2\simeq 1.3
\label{protonneutronratio}
\end{equation}
(see \cite{ericson88} p. 276)
the recoil terms $E^{(0){\rm Born}}_{0+}$ may be
replaced by multiplying $E^{(-){\rm Born}}_{0+}$
with $(1+\frac{m_\pi}{m})^{-1/2}$
and $(1+\frac{m_\pi}{m})^{+1/2}$ in order to get the results for the 
proton and neutron, respectively. The relation given in 
(\ref{protonneutronratio}) is well justified at threshold but its approximate
validity extends to higher energies \cite{hanstein98,drechsel99,donnachie72}.
The pseudovector coupling constant $f$ in (\ref{E0+}) is given by
$f=g_{\pi NN}(m_\pi/2m)$ with $g_{\pi NN}= 13.169\pm 0.057$.
There is a remarkable agreement between the numbers given in Table 
\ref{table2} but these numbers are larger by a factor $\sim 2.4$
than the corresponding numbers in line 9 of Table \ref{drechsel}. Two reasons 
for the deviation of the empirical $E_{0+}$ amplitude 
from the Born approximation
have been discussed in \cite{drechsel99}. The first reason is that the
pseudovector (PV)  coupling  is not valid at high photon energies but has to be
replaced by some average of the  PV and pseudoscalar the  (PS) coupling.
The second reason are $\rho$ and $\omega$ meson $t$-channel exchanges
which are not taken into account in the Born approximation.

In Table \ref{drechsel} (see also Table \ref{armstrong}) 
we see that the different resonant contributions to the
electric polarizabilities
cancel each other, so that the electric polarizabilities are mainly due to the 
$t$-channel part $\alpha^t_{p,n}$ ($\sim$60\%) given in line 8
and a smaller nonresonant part
$\alpha(E_{0+})$ ($\sim$30\%) given in line 9. 
For the magnetic polarizabilities  there is an
almost complete cancellation of the $P_{11}(1440)$, $D_{13}(1520)$ 
and $F_{15}(1680)$
contributions, so that the main remaining 
contributions are due to the  $P_{33}(1232)$
resonance, canceled to a large extent by  the $t$-channel contribution
$\beta^t$. The  nonresonant background given in line 10
of Table \ref{drechsel} amounts to about  $10$\% of the experimental 
electric polarizabilities
and to about 70\% of the experimental magnetic polarizabilities. 
This means that precise predictions of these contributions are highly
desirable, especially for the magnetic polarizabilities.  
Unfortunately, the
non-$E_{0+}$ parts of the nonresonant photoabsorption cross sections are
dominated by two-pion channels where the information on the multipole  content
is scarce.
\subsection{Discussion of the $t$-channel contribution} 

In \cite{schumacher06} it has been shown that there are two independent,
but apparently equivalent and complementary options to calculate the 
scalar-isoscalar 
$t$-channel contribution to the electromagnetic polarizabilities
of the nucleon. 

Option 1 makes use of the properties of the $\sigma$-meson as
predicted by the quark-level NJL model and in this respect is of course
model dependent. The quark-level  NJL model predicts a definite 
$\sigma$-mesons mass, {\it viz.} $m_\sigma=666$ MeV, through a parameter-free
relation of $m_\sigma$ to the pion decay constant $f_\pi$. The result
$(\alpha-\beta)^t=15.2$ is in an excellent agreement with the experimental
result. The agreement between a prediction and an experimental result
cannot be used as an argument for the validity of the prediction without
further support. This support is provided by dispersion theory applied to the
measured properties of the $\sigma$ meson as showing up in particle
reactions with two pions in the intermediate state (Option 2).

Option 2 first takes into consideration that the $\sigma$ meson
has been  observed in  many data analyses \cite{yao06} as a pole on the 
second sheet  of the isoscalar $S$ wave of $\pi\pi$ scattering. 
This pole describes 
part of the resonant structure of the $\sigma$ meson without being a complete
description. This latter property of the pole follows from the fact that the
$90^\circ$ crossing of the scalar-isoscalar phase $\delta^0_0(s)$ 
is located at much higher energies than  predicted by the
structure of the pole. The analyses of Colangelo et al. \cite{colangelo01}
and Caprini et al. \cite{caprini06}
led to 
\begin{eqnarray}
&&\sqrt{s}{\rm (pole)} = (470\pm 30) - i(295\pm 20)\,\, {\rm MeV}\quad 
\sqrt{s}(\delta_S=90^\circ)=(844\pm 13)\,\, {\rm MeV}
\mbox{\cite{colangelo01}}, 
\label{colangelo}\\
&&M_\sigma=441^{+16}_{-8}\,\,{\rm MeV},\quad 
\Gamma_\sigma=544^{+18}_{-25}\,\,{\rm MeV} \mbox{\cite{caprini06}}.
\label{caprini}
\end{eqnarray}
The numbers contained in    (\ref{colangelo}) and (\ref{caprini}) are
extremely valuable in characterizing the properties of the $\sigma$ meson
as a real particle but they can only qualitatively  be compared with the mass
$m_\sigma=666$ MeV of the virtual $\sigma$ meson, because in the latter
case there is no open decay channel. This means that there is no contradiction 
between the existence of the broad mass distribution for the real $\sigma$
meson and a precisely determined mass of the virtual $\sigma$ meson.  
Furthermore, the numbers   contained in  (\ref{colangelo}) and (\ref{caprini})
are of no direct  relevance for the prediction of $(\alpha-\beta)^t$. First
of all it certainly would lead only to a qualitative estimate for
$(\alpha-\beta)^t$ if the parameters of the $\sigma$-meson pole in
(\ref{colangelo}) and (\ref{caprini}) would be used instead of $m_\sigma=666$
MeV. Furthermore, such an insufficient attempt is not necessary because the
BEFT \cite{bernabeu74} sum rule provides a precise relation between
$(\alpha-\beta)^t$ and the properties of the real $\sigma$ meson. In the BEFT
sum rule the imaginary part of the $t$-channel  Compton scattering
amplitude is given by an unitarity relation where the two reaction
$\gamma\gamma\to\sigma\to\pi\pi$  and $N\bar{N}\to\sigma\to\pi\pi$ are
exploited. In these reactions the resonant structure of the $\sigma$ meson
enters via the experimentally determined scalar-isoscalar phase 
$\delta^0_0(s)$ which is considerably  different from the corresponding
quantity predicted by the poles shown in  (\ref{colangelo}) and
(\ref{caprini}). The real part of the $t$-channel Compton
scattering amplitude is obtained 
via a dispersion relation. The present status of the evaluation of the BEFT
sum rule $(\alpha-\beta)^t_{np}=15.3\pm 1.3$ is in good agreement with the 
experimental result as well as the prediction based on the quark-level
NJL model.

\section{Conclusion}
The good agreement of the result based on  the BEFT sum rule with 
the experimental result as well as the prediction based on the quark level
NJL model may be understood as a strong argument that the two 
predictions of $(\alpha-\beta)^t$ are equivalent.  This implies that in
addition to the poles in (\ref{colangelo}) and (\ref{caprini}) also the mass
$m_\sigma=666$ MeV of the virtual $\sigma$ meson 
is an experimentally verified property of
the $\sigma$ meson.

\section*{Acknowledgment}
The author is indebted to Deutsche Forschungsgemeinschaft for the 
support of this
work through the projects SCHU222 and 436RUS113/510. He thanks M.I. Levchuk,
A.I. L'vov and A.I. Milstein for a long term cooperation which contributed 
to the motivation for   the present investigation.

\end{document}